\documentclass{PoS}

\usepackage{lscape}
\usepackage{afterpage}
\usepackage{amsmath}
\usepackage[square,comma,numbers,sort&compress]{natbib}
\usepackage{float}

\newcommand*{\La}{\cal{L}}
\newcommand*{\no}{\noindent}
\newcommand*{\bea}{\begin{eqnarray}}
\newcommand*{\eea}{\end{eqnarray}}
\newcommand*{\be}{\begin{equation}}
\newcommand*{\ee}{\end{equation}}
\newcommand*{\pd}{\partial}
\newcommand*{\pdm}{\pd_{\mu}}
\newcommand*{\pdn}{\pd_{\nu}}

\newcommand*{\pref}[1]{(\ref{#1})}

\newcommand*{\mn}{{\mu\nu}}

\newcommand*{\nn}{\nonumber}

\newcommand{\bma}{\begin{pmatrix}}
\newcommand{\ema}{\end{pmatrix}}
\newcommand*{\la}{\left\langle}
\newcommand*{\ra}{\right\rangle}

\title{Testing gauge-invariant perturbation theory\footnote{Combined proceedings of the talks of both authors.}}

\ShortTitle{Gauge-invariant perturbation theory}

\author{\speaker{Pascal T\"orek}\thanks{Supported by the FWF DK W1203-N16}\\
        E-mail: \email{pascal.toerek@uni-graz.at}}

\author{\speaker{Axel Maas}\\
        E-mail: \email{axel.maas@uni-graz.at}\\\\
        Institute of Physics, NAWI Graz, University of Graz, Universit\"atsplatz 5, 8010 Graz, Austria}

\abstract{Gauge-invariant perturbation theory for theories with a Brout-Englert-Higgs effect, as developed by Fr\"ohlich, Morchio and Strocchi, starts out from physical, exactly gauge-invariant quantities as initial and final states. These are composite operators, and can thus be considered as bound states. In case of the standard model, this reduces almost entirely to conventional perturbation theory. This explains the success of conventional perturbation theory for the standard model. However, this is due to the special structure of the standard model, and it is not guaranteed to be the case for other theories.

Here, we review gauge-invariant perturbation theory. Especially, we show how it can be applied and that it is little more complicated than conventional perturbation theory, and that it is often possible to utilize existing results of conventional perturbation theory. 

Finally, we present tests of the predictions of gauge-invariant perturbation theory, using lattice gauge theory, in three different settings. In one case, the results coincide with conventional perturbation theory and with the lattice results. In a second case, it appears that the results of gauge-invariant perturbation theory agree with the lattice, but differ from conventional perturbation theory. In the third case both approaches fail due to quantum fluctuations. }

\FullConference{34th annual International Symposium on Lattice Field Theory\\
		24-30 July 2016\\
		University of Southampton, UK}

\begin{document}

\section{Introduction}

A fundamental requirement of gauge theories is that all physical, observable quantities have to be gauge-invariant \cite{Banks:1979fi,'tHooft:1979bj,Frohlich:1980gj}. Especially, all experimentally observable particles have to be gauge-invariant states. In non-Abelian gauge theories, any operator describing such gauge-invariant states is necessarily composite. The states described by composite operators can be dubbed, in analogy to QCD, bound states. In contradistinction, the elementary fields are gauge-dependent, and therefore cannot describe physical degrees of freedom.

In the electroweak sector of the standard model this leads to an apparent paradox \cite{Banks:1979fi,'tHooft:1979bj,Frohlich:1980gj}: Its description, including experimental results, is in terms of the elementary Higgs and $W$ and $Z$ particles, as well as the elementary fermions \cite{pdg,Bohm:2001yx}, and very successfully so. In fact, these gauge-dependent particles appear to be observable states, and their perturbative description \cite{Bohm:2001yx} is extremely accurate.

The resolution of this paradox is that the physical states are still gauge-invariant composite states. However, their properties coincide essentially with those of the elementary particles as far as possible. This can be seen using an approach \cite{Frohlich:1980gj,Frohlich:1981yi} called gauge-invariant perturbation theory\footnote{Note, as it will become clear below, that this is different from the gauge-invariant perturbation theory of  \cite{Langguth:1985eu,Philipsen:1996af}, which is slightly deforming the theory \cite{Maas:2013aia} and cannot be applied in the presence of QED or Yukawa couplings to fermions.} \cite{Seiler:2015rwa}. In this approach it is explicitly seen how the description of the bound state dynamics reduces to conventional perturbation theory \cite{Frohlich:1980gj,Frohlich:1981yi}. While this appears to be rather odd at first glance, this result has been confirmed in lattice calculations \cite{Maas:2012tj,Maas:2013aia}. However, this works because of the very special structure of the standard model \cite{Maas:2015gma}, and indeed seems not to be holding in general \cite{Maas:2016ngo}. Also, as a perturbative description, this equivalence may break down for some range of the parameters of the theory \cite{Maas:2013aia}.

All of this will be reviewed in the following. The basic outline of the gauge-invariant perturbation theory of \cite{Frohlich:1980gj,Frohlich:1981yi} will be described in Section \ref{s:gipt}. Then two classes of theories will be discussed in turn. One is where for structural reasons it is possible that gauge-invariant perturbation theory can coincide with conventional perturbation theory, but can still be limited by quantum fluctuations, in Section \ref{s:sm}. A case where it is even structurally impossible will be discussed in Section \ref{s:gut}. In both cases the predictions of both kinds of perturbation theory will be compared to lattice results. It will be seen that if perturbative calculations are possible at all, then gauge-invariant perturbation theory always provides results which coincide with the lattice results, but conventional perturbation theory does not. However, as will be seen in Section \ref{s:gipt}, the effort for gauge-invariant perturbation theory is often only slightly larger than for conventional perturbation theory, and therefore this does not imply too much additional complications.

These findings will be wrapped up in Section \ref{s:conclusion}.

\section{Gauge-invariant perturbation theory}\label{s:gipt}

Gauge-invariant perturbation theory, based on the Fr\"ohlich, Morchio and Strocchi (FMS) mechanism \cite{Frohlich:1980gj,Frohlich:1981yi}, requires a theory in which a Brout-Englert-Higgs effect \cite{Bohm:2001yx} is at work.

In the following, the comparatively simple case of the propagation of a single particle will be treated. The extension to scattering processes will be studied elsewhere \cite{Egger:unpublished}, but works along essentially the same lines \cite{Maas:2012tj}.

The starting point is a theory which includes a Higgs field, in some representation of the gauge group:
\bea
{\La}&=&-\frac{1}{4}W_\mn^aW^\mn_a+(D_\mu\phi)^\dagger D^\mu\phi-\gamma\left(\phi^\dagger\phi-v^2\right)^2\;,\nn\\
W_\mn^a&=&\pdm W_\nu^a-\pdn W_\mu^a-g f^{abc} W_\mu^b W_\nu^c \;,\nn\\
D_\mu^{ij}&=&\pd_\mu\delta^{ij}-igW_\mu^a\tau^{ij}_a\;,\nn
\eea
\no with the Higgs field $\phi$, the gauge field $W$, the gauge coupling $g$ and the parameters of the Higgs potential $m_0$ and $\gamma$. The $f$ are the usual structure constants, and the $\tau$ are the relevant matrices for the representation of the Higgs field. The parameters of the model should be such that the Brout-Englert-Higgs (BEH) effect is active\footnote{This is a non-trivial statement, and will be discussed in more detail in Section \ref{s:sm}.}. The theory may furthermore furnish some global, i.\ e.\ custodial symmetry. The addition of fermions and QED is straightforward \cite{Frohlich:1980gj,Frohlich:1981yi,Maas:2016qpu,Egger:unpublished}, but will be skipped for simplicity here.

Any physical state must now be gauge-invariant \cite{Banks:1979fi,'tHooft:1979bj,Frohlich:1980gj}. Thus, only spin, parity, and other global quantum numbers can characterize a state.

Consider now the simplest possibility, a scalar state, which is otherwise a singlet. The simplest gauge-invariant operator in this channel is ${\cal O}_{0^+}(x)=\phi_i^\dagger(x)\phi_i(x)$. The simplest particle having these quantum numbers therefore has the propagator
\be
\la{\cal O}_{0^+}^\dagger(x){\cal O}_{0^+}(y)\ra\label{zp}\;.
\ee
\no Gauge-invariant perturbation theory \cite{Frohlich:1980gj,Frohlich:1981yi} now proceeds as follows to determine the properties of these states: First, choose a gauge in which the Higgs vacuum expectation value $nv$ does not vanish, e.\ g.\ 't Hooft gauge \cite{Bohm:2001yx}, where $v$ is the absolute value and $n$ is the direction in the gauge group. It will be assumed that there is only a single (degenerate set of) minima in the potential.
Then rewrite the Higgs fields in \pref{zp} as $\phi_i(x)=vn_i+\eta_i(x)$ with $\la\eta_i(x)\ra=0$, yielding
\bea
\la{\cal O}_{0^+}^\dagger (x){\cal O}_{0^+}(y)\ra = 
&&d + 
4v^2~\la \Re\left[n_i^\dagger\eta_i\right]^\dagger(x)~\Re\left[n_j^\dagger\eta_j\right](y) \ra  \label{gipt}\\
+&&2v\left(\la (\eta_i^\dagger\eta_i)(x)~\Re\left[n_j^\dagger\eta_j\right](y) \ra + (x\leftrightarrow y) \right) + 
\la (\eta_i^\dagger\eta_i)(x)~(\eta_j^\dagger\eta_j)(y) \ra  \;,\nn
\eea
\no where terms with vanishing expectation values have been dropped and $d$ collects various space-time-independent constants. None of the terms on the right-hand side of \pref{gipt} is individually gauge-invariant, but their sum is. 

The next step is to expand the right-hand-side in the conventional perturbative series. This yields to leading order
\bea
\la{\cal O}_{0^+}^\dagger(x){\cal O}_{0^+}(y)\ra& =& d'+4v^2\la \Re\left[n_i^\dagger\eta_i\right]^\dagger(x)~\Re\left[n_j^\dagger\eta_j\right](y) \ra_\text{tl}\nn \\
&&+\la \Re\left[n_i^\dagger\eta_i\right]^\dagger(x)~\Re\left[n_j^\dagger\eta_j\right](y) \ra_\text{tl}^2+{\cal O}(g^2,\gamma)\nn\;,
\eea
\no where 'tl' denotes the tree-level value, and $d'$ is again a collection of space-time-independent constants. These constants are irrelevant. The term proportional to $4v^2$ describes the propagation of a single Higgs particle. The last term describes two propagating non-interacting Higgs particles, starting and terminating at the same points in space and time. It is obtained by the fact that at tree-level the full four-point function decomposes by cluster decomposition into a product of two tree-level Higgs propagators, as there is no connected vertex function at order $\gamma^0$. This result implies that the right-hand side has two poles, one at the (tree-level) Higgs mass and one at twice the (tree-level) Higgs mass. Comparing poles on both sides, gauge-invariant perturbation theory predicts that the physical, gauge-invariant left-hand side should have a mass equal to the tree-level mass of the Higgs and the next state should then be a scattering state of twice the same particle. Thus, as announced, the bound states have the same properties as the gauge-dependent elementary states, especially the same mass\footnote{In fact, the statement is even stronger: To this order the left-hand side and the right-hand side should coincide in total.}.

Higher-order corrections can be included by going to higher orders in conventional perturbation theory on the right-hand side. Since the mass of the Higgs then becomes scheme-dependent \cite{Bohm:2001yx}, this requires a suitable scheme, like the pole scheme. In addition, further scattering poles involving more particles will appear in this way.

Of course, as the full correlation functions appear on the right-hand side in \pref{gipt}, there can, in principle, also be non-perturbative corrections adding further poles. If these corrections are small, and no further poles are created, then the results coincide to a good accuracy with conventional perturbation theory. This explains the success of conventional perturbation theory, if no such additional non-perturbative poles and contributions are substantial in any channel. For the standard model, this seems to be the case \cite{pdg}. However, non-perturbative contributions are not necessarily zero, and possible consequences of them will be studied elsewhere \cite{Egger:unpublished}. For the following it will be assumed that they can be neglected. Furthermore, the appearance of the full three-point and four-point functions adds also further perturbative contributions, which may not be irrelevant. Also this issue will be discussed in more detail elsewhere \cite{Egger:unpublished}.

The same recipe applied above to the propagation of a single particle can also be applied to any other process. In particular, this can be done for scattering processes, and reproduces in the standard model conventional perturbation theory to the appropriate order \cite{Egger:unpublished,Maas:2012ct}.

The procedure thereby outlined is the basic principle of gauge-invariant perturbation theory \cite{Frohlich:1980gj,Frohlich:1981yi}:
\begin{itemize}
 \item Choose an operator with the desired quantum numbers and write down a gauge-invariant operator carrying these quantum numbers.
 \item Rewrite the involved Higgs fields by splitting off the Higgs vacuum expectation value.
 \item Expand the correlation functions perturbatively.
\end{itemize}
Because the appearing correlation functions have to carry the same spin and parity it follows that two-point correlation functions can at most appear if an elementary particle of the same spin and parity exists. Thus, only at most in these cases a single-particle pole can appear, while for all other spin and parity channels only scattering states can surface.

Furthermore, in principle all operators carrying the same quantum numbers can mix in. Considering multiple operators, individual contributions can be isolated by diagonalization. However, in general it cannot be expected that it will be possible to diagonalize both sides simultaneously.

The situation becomes more involved if the gauge-invariant operator also carries global quantum numbers. If the corresponding symmetry is unbroken, then according to the Wigner-Eckhart theorem the gauge-invariant operator should exhibit a degeneracy in accordance with its multiplet structure. The appearance of such degeneracies is important for being able to replicate the gauge multiplet structure.

Arguably the most important example is the standard model case with its SU(2) local and global custodial symmetry. Neglecting QED the weak gauge bosons form a gauge triplet. A corresponding gauge-invariant operator is given by a custodial triplet vector, and its correlator expands to
\be
\langle(\tau^a_{ij}\varphi_j^\dagger D_\mu\varphi_i)(x)(\tau^a_{kl}\varphi_k^\dagger D_\mu\varphi_l)(y)\rangle=
c\langle W^b_\mu(x) W^b_\mu(y)\rangle+{\cal O}(\eta)\label{wcor}\;,
\ee
\no where the $\tau$ are generators of the custodial symmetry, $\varphi$ is the angular part of the Higgs field \cite{Maas:2013aia}, and $c$ is some constant. By the same argument as above, the custodial triplet has the same mass as the $W$ bosons in the corresponding gauge. In this case, the Nielsen identities \cite{Nielsen:1975fs} make the comparison simpler beyond tree-level. Hence, the multiplet structure of the gauge symmetry is exchanged for a custodial multiplet structure.

This immediately leads to the question of what happens in situations where such a matching is not possible \cite{Maas:2015gma}. An example of this problem, to be treated in more detail in Section \ref{s:gut}, is a theory with gauge group SU(3) and a single Higgs field in the fundamental representations, and thus only a U(1) custodial symmetry. In this case, the simplest vector operator is a custodial singlet, and expands as \cite{Maas:2016ngo}
\bea
\langle O_\mu(x) O^\dagger_\mu(y)\rangle&=&v^4\langle W_\mu^8(x)W_\mu^8(y)\rangle + \mathcal{O}(\eta)\label{wcor2}\;,\\
O_\mu(x)&=&i(\phi^\dagger D_\mu \phi)(x)\;,\label{wcor2a}
\eea
\no and thus has the same mass as one of the gauge bosons only, and consequently only has a single state. The mass of this state is actually the heaviest of the perturbative spectrum \cite{Maas:2016qpu}. Of course, the determination on the right-hand side can then use again conventional perturbation theory. It is possible to do the same in other quantum channels, and also for other theories. Further examples can be found in \cite{Torek:2015ssa,Maas:2016qpu,Maas:2015gma,Maas:2016ngo}.

This completes the description of gauge-invariant perturbation theory. As will be seen in sections \ref{s:sm} and \ref{s:gut}, gauge-invariant perturbation theory indeed gives the correct description of the spectrum. As the last example shows, this can make a relevant difference in the spectrum to be expected in experiment. This could even rule out or include new proposals of physics beyond the standard model \cite{Maas:2015gma}.

The usage of gauge-invariant perturbation theory requires only little more effort than conventional perturbation theory. In addition, results of conventional perturbation theory can be used to obtain results for gauge-invariant perturbation theory, and therefore previous results are actually very valuable. It appears reasonable to employ it instead of conventional perturbation theory for theories which do show a BEH effect, provided they can be treated perturbatively at all.

\section{The standard model case}\label{s:sm}

While gauge-invariant perturbation theory is rather straightforward to apply, confirming its claims is not entirely trivial. For the standard model case, it reduces essentially to conventional perturbation theory, and therefore it is hard, if at all possible, to find differences in experimental data \cite{Egger:unpublished}. On the other hand, the operators to be described are bound state operators, and therefore non-perturbative methods are required to determine their spectrum, which will be done here using lattice methods. Most interesting are those cases in which the predictions of gauge-invariant perturbation theory differ from those of conventional perturbation theory. This will be investigated in Section \ref{s:gut}. But before this, it is necessary to establish as a baseline that it works for the standard model, though the agreement with experiment can already be taken as a strong indication that it does.

Unfortunately, simulating the full standard model is currently out of reach of lattice calculations. The most important part is, however, the weak sector containing the Higgs and the $W$ and $Z$ bosons, which is readily accessible in lattice simulations. Here, the results of \cite{Maas:2012tj,Maas:2013aia,Maas:2014pba} will be summarized and extended with some new data. Also all the details of the simulations can be found there.

Gauge-invariant perturbation theory requires a BEH effect to be applicable. That is actually not a trivial requirement, as the presence of the BEH effect is depending on the gauge, as has been explicitly verified on the lattice \cite{Caudy:2007sf}, and as follows from the Osterwalder-Seiler-Fradkin-Shenker argumentation \cite{Osterwalder:1977pc,Fradkin:1978dv}. But it is actually sufficient that in the gauge chosen the BEH effect is at work \cite{Maas:2013aia}.

\begin{figure}
\centering
\includegraphics[width=\linewidth]{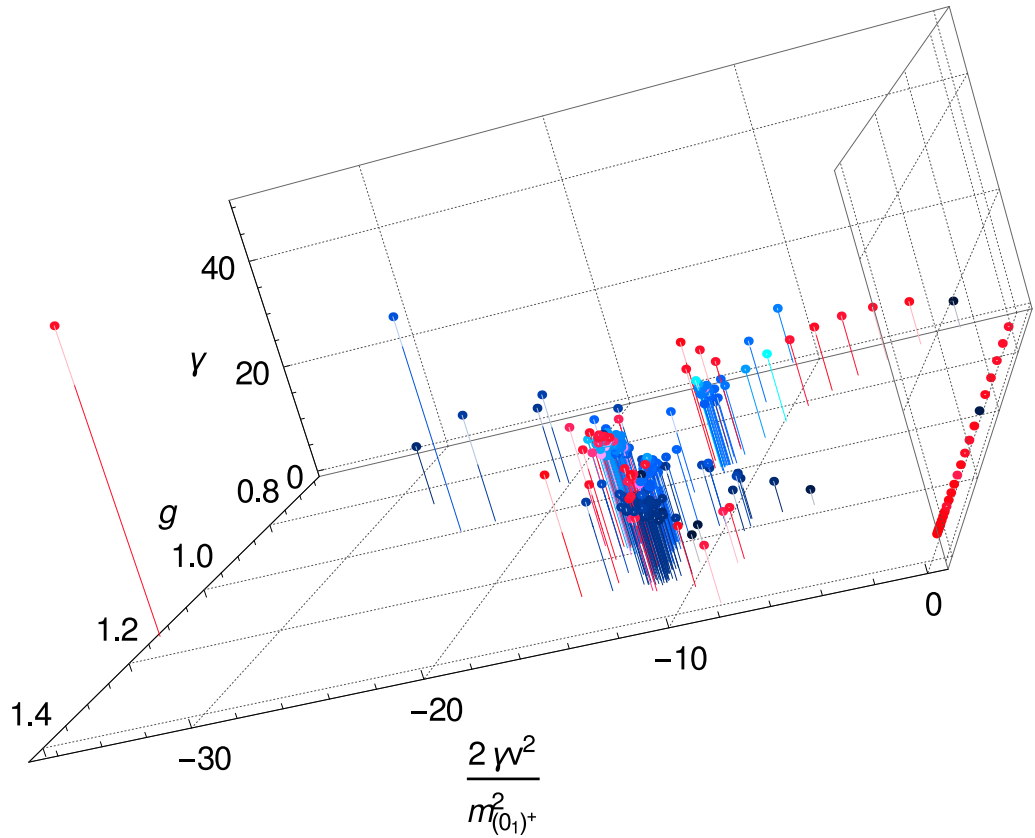}
\caption{\label{fig:lcp}The phase diagram of the weak sector in terms of the continuum parameters. The dimensionful parameter $v$ is measured in units of the mass of the ground state in the $0^+$ channel, see \cite{Maas:2014pba}. The figure includes the data of \cite{Maas:2014pba} and further additional points. Red points have in the aligned Landau gauge \cite{Maas:2012ct} no BEH effect while blue points do. The lighter the point the finer is the lattice in units of the custodial triplet vector mass.}
\end{figure}

However, as has already been observed in the first lattice calculations of this theory \cite{Evertz:1985fc,Langguth:1985dr}, it is not sufficient that a BEH effect is possible at tree-level. Large quantum fluctuations can drive the system outside the BEH regime. The opposite effect has not been observed so far. The corresponding phase diagram is shown in Figure \ref{fig:lcp}. It is visible that a too large Higgs self-interaction and a too shallow classical potential will drive the system out of the BEH regime, while the gauge coupling makes only at very strong coupling a difference. In the region without BEH effect gauge-invariant perturbation theory is not applicable. However, in this region the physics is also QCD-like \cite{Maas:2013aia}, and therefore also conventional perturbation theory cannot be used to determine the gauge-invariant spectrum of the theory, just as in QCD.

\begin{figure}
\centering
\includegraphics[width=\linewidth]{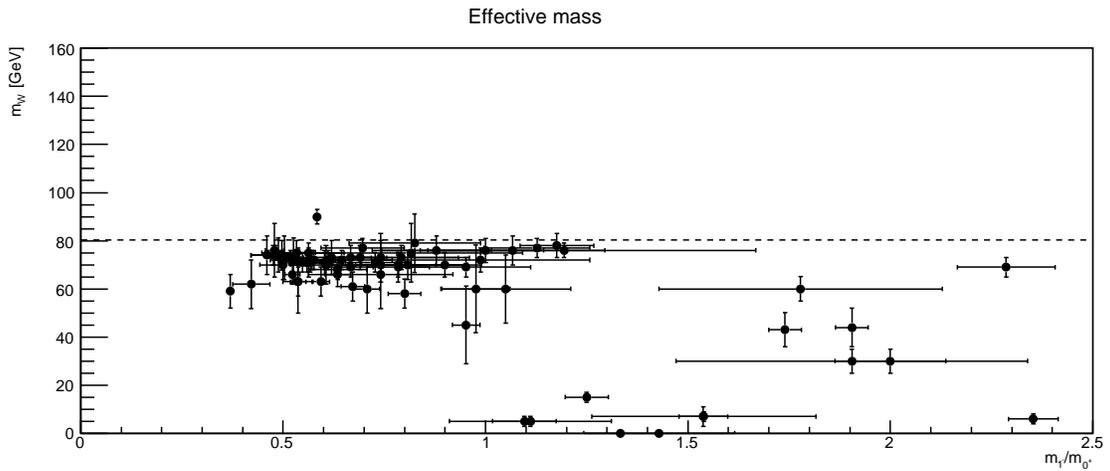}
\caption{\label{fig:w}The mass of the $W$/$Z$ gauge boson as a function of the ratio of the masses of the ground states of the custodial vector triplet state to the one of the custodial singlet scalar. This plot combines new data with the one from \cite{Maas:2013aia}. The results are not extrapolated to infinite volume.}
\end{figure}

To test gauge-invariant perturbation theory the comparison of the pole mass on both sides of \pref{wcor} is probably the most interesting quantity, as here no renormalization issues play a role. An investigation of the scalar singlet case can be found in \cite{Maas:2013aia}. Though more subtle, the results are in agreement with those of gauge-invariant perturbation theory to the extent possible.

The result is shown in Figure \ref{fig:w}, as a function of the ratio of the ground state masses of the custodial triplet vector state and the custodial singlet scalar, i.\ e.\ the physical equivalences of the conventional $W$/$Z$ gauge bosons and the Higgs. If the ratio is between 1/2 and 1, the results almost agree with the 80 GeV prediction of gauge-invariant perturbation theory. The fact that they are slightly too small is a finite-volume effect, which has not yet been corrected for \cite{Maas:2013aia}.

Below a ratio of 1/2 the elastic decay threshold opens for the scalar, and a suitable precise determination of resonances was not yet possible, if they exist at all \cite{Maas:2014pba}.

If the ratio exceeds 1, something interesting happens: The $W$-mass becomes substantially lower than the predictions. In fact, a detailed analysis shows \cite{Maas:2013aia} that only at very short distances any notion of a mass is actually applicable, and the long-range properties become very much QCD-like \cite{Maas:2011se}, and there is likely not even a reasonable pole anymore present. In fact, if the ratio exceeds 1 then the system actually behaves completely QCD-like, i.\ e.\ in this gauge no longer any BEH effect is observed \cite{Maas:2013aia,Maas:2014pba}, an effect already indicated in much earlier calculations \cite{Evertz:1985fc}. Though this is, of course, a purely numerical observation, this appears to indicate that a BEH effect in this theory is not possible for the lightest state being the scalar rather than the vector.  In the full standard model this may be influenced by contributions from the matter sector, but it is already very odd that it is not possible in this theory: According to (gauge-invariant or not) perturbative calculations, this should be possible. This could indicate another limitation due to quantum fluctuations, though its source and detailed mechanism is yet unknown.

In total, the investigation of this standard-model-like case shows that gauge-invariant perturbation theory correctly predicts the physical spectrum, though this result does not deviate from those of conventional perturbation theory. On the other hand, the results show that arguments like sufficiently weak couplings or the existence of a BEH effect at tree-level are not enough to ensure that (either) perturbative description works. The mysterious lower mass limit for the scalar is also a strong indication that applicability of perturbation theory of either type may be harder to predict than originally anticipated, an issue which may also affect theories beyond the standard model \cite{Maas:2015gma}.

\section{A structural mismatch}\label{s:gut}

In the standard model case the multiplet structure was such that it was possible to map the results of gauge-invariant perturbation theory to conventional perturbation theory. A situation, where this is not possible is the following \cite{Torek:2015ssa,Maas:2016ngo}: Consider an SU(3) gauge group with a single flavor of Higgs particles in the fundamental representation. A BEH effect is then possible in the same sense as for the standard model, though quantitatively again tree-level calculations do not sufficiently reliable indicate where in the phase diagram \cite{Maas:2016ngo}.

It is straightforward \cite{Bohm:2001yx,Torek:2015ssa} to determine the tree-level mass spectrum in conventional perturbation theory. There is a massive scalar, and five out of the eight gauge bosons become massive. The three massless ones correspond to the ones of the unbroken SU(2) subgroup. The five heavier ones split into a quadruplet of degenerate lighter ones and a single more heavier one.

The only global symmetry this theory has, up to some discrete symmetries, is an additional U(1) \cite{Maas:2016ngo}. This symmetry is carried again by the Higgs field, and acts similar to a baryon number. In fact, this theory is without BEH effect nothing but scalar QCD with one flavor.

This implies that the physical spectrum can be separated, for any spin, parity and charge-parity, in superselection sectors of this U(1) charge. Since the Higgs particles are bosons, each charge superselection sector contains only bosons, of any integer spin. A-priori, based on the Wigner-Eckhart theorem, it would be expected that none of these states are degenerate barring accidental degeneracies. Thus, there is no natural candidate to describe the degenerate vector states, while the single scalar particle is easier to assign. 

However, neglecting for a moment the problem of degeneracies a second problem arises. Any gauge-invariant state can, as in QCD, only carry any multiple of three of the Higgs U(1) charges. Again, just like baryon number. Therefore, to map successfully any states at all in gauge-invariant perturbation theory it must be allowed for a mismatch in the U(1) charge by a factor of three.

Accepting this, it turns out that actually an uncharged scalar, in the same way as in \pref{gipt}, has the same mass as the Higgs particle \cite{Torek:2015ssa}. It is therefore not protected by the U(1) charge against decays, only by kinematic reasons. On the other hand, the lightest state with non-vanishing U(1) charge must necessarily be stable. But so far we have not been able to identify an operator with non-vanishing U(1) charge which expands in gauge-invariant perturbation theory to a single particle pole. All of those investigated so far only created scattering states. However, this may still be resolved.

More interesting is again the question of the vector particles. To have similar stability features as the elementary gauge bosons it appears reasonable to look at vector particles with vanishing U(1) charge. The result \pref{wcor2} indicate, however, that according to gauge-invariant perturbation theory this channel should be gapped, and the lightest state should have the same mass as the heaviest of the gauge bosons.

\begin{figure}
\centering
\includegraphics[width=0.49\textwidth]{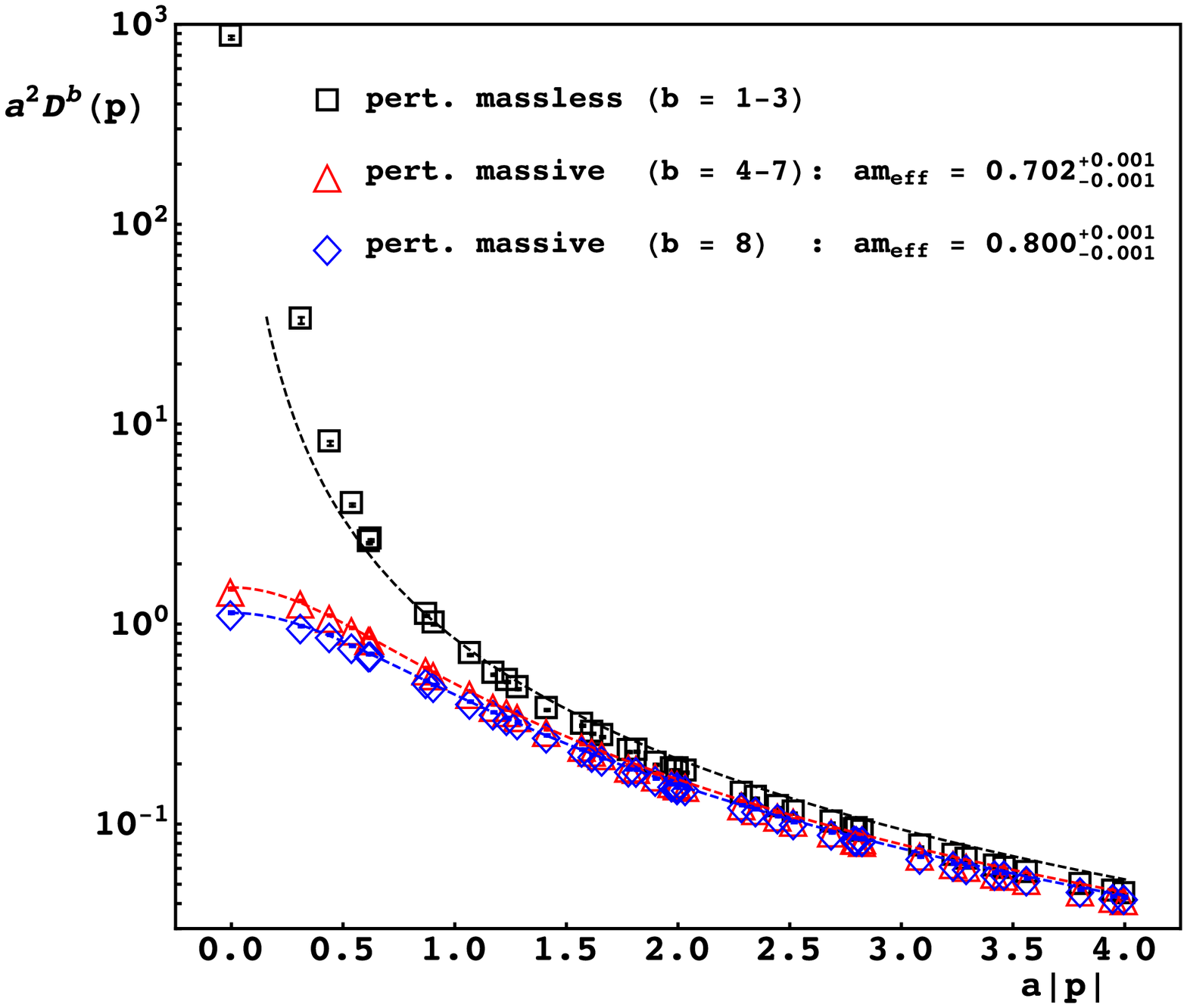}
\includegraphics[width=0.50\textwidth]{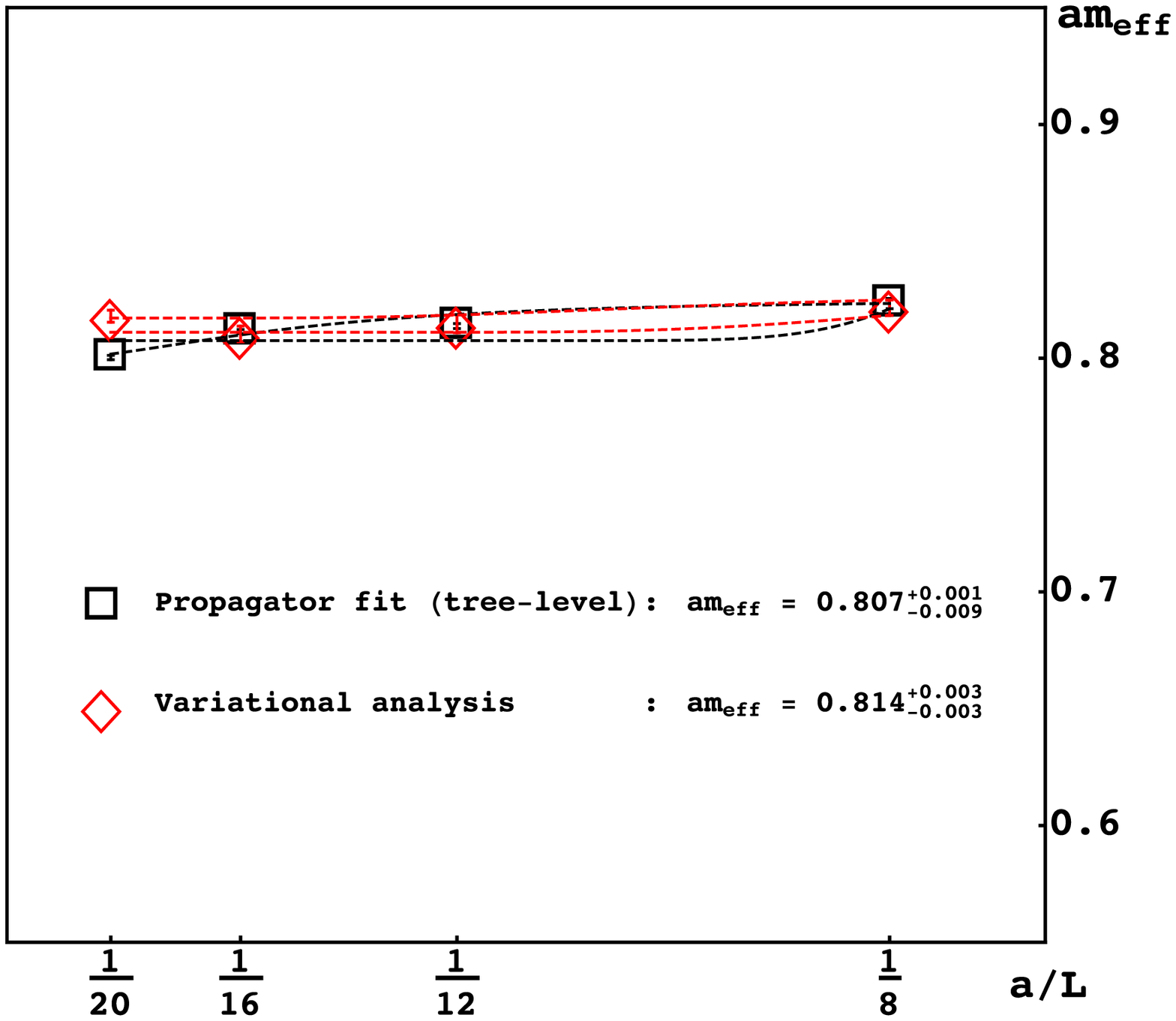}
\caption{\label{fig:gut} In the left panel the aligned Landau gauge \cite{Maas:2012ct} results for the propagators of the gauge bosons in momentum space are shown, together with tree-level fits. On the right-hand side the masses obtained from these fits are compared to the mass extracted from the physical state \protect\pref{wcor2a}. Data are from \cite{Maas:2016ngo}.}
\end{figure}

We have studied therefore this channel on the lattice \cite{Maas:2016ngo}. The results are shown in Figure \ref{fig:gut} for a particular point in the phase diagram where the BEH effect is active \cite{Maas:2016ngo}. Multiple observations can be made.

The first is that the gauge boson propagators are very close to tree-level, and thus massive or massless depending on whether they are in the unbroken subgroup or in the coset. Also, the degeneracies coincide with perturbation theory, though this is not shown explicitly. Finally, the mass ratio of the two appearing masses is close to its tree-level value, though deviates slightly \cite{Maas:2016ngo}. This very good agreement with leading order conventional perturbation theory strongly suggests that the physics at these parameters is indeed essentially perturbative.

The right-hand side plot shows then a comparison of the heavier of the masses extracted from the propagator compared to the lowest mass obtained from the gauge-invariant, physical correlator \pref{wcor2a}, as a function of the volume. There appears to be only a single, non-degenerate ground state in this channel, as expected from the above arguments. The other states, which are not shown, in this channel appear to be much heavier, probably being at or above the expected elastic threshold \cite{Maas:2016ngo}. Moreover, the ground-state mass shows essentially no dependence on the volume and a substantially non-zero mass for all volumes. This strongly suggests that this state is indeed massive, and the channel is gapped. Thus, no trace of the massless states are seen.

In fact, just from considerations of the gauge-invariant part of the physics, there is no obvious argument why this theory should not have a mass gap. Although the Osterwalder-Seiler-Fradkin-Shenker argument \cite{Osterwalder:1977pc,Fradkin:1978dv} is not applicable in this case, as the gauge symmetry is not entirely hidden by the Higgs, its main conclusion still holds: There is no fundamental distinction between the BEH effect and confinement at the gauge-invariant level. Given that confinement induces a mass gap it appears therefore natural that this theory also should have a mass-gap throughout the phase diagram. This conclusion is rather intriguing, but at the current time more speculation than conjecture.

Returning to the predictions of gauge-invariant perturbation theory, the figure gives a much more clean-cut result: The mass as obtained from the propagator fit is in good agreement with the one form the gauge-invariant channel, as predicted from \pref{wcor2a}. This supports the approach of gauge-invariant perturbation theory.

Of course, this is not a proof. Especially, there are deviations in the fits, which is to be expected for a tree-level fit, especially for the massless elementary propagator. There are also several possibilities for systematic errors due to lattice calculations \cite{Maas:2016ngo}, especially in the choice of operator basis. However, none of these artifacts strongly biases the result towards an agreement with gauge-invariant perturbation theory. Especially, discretization and finite volume effects should make the presence of heavier or lighter modes more visible, and especially push the result away from agreement. Also, the operator itself has the same, i.\ e.\ zero, overlap with any of the gauge-dependent fields. Still, more elaborate investigations are necessary and will be done \cite{Maas:unpublishedtoerek}.

\section{Conclusion}\label{s:conclusion}

Gauge-invariant perturbation theory is a logically well-funded concept in field theory. It is completely consistent with the success of conventional perturbation theory in the standard model, and at the same time indicates improvement potential. At the same time, in the wider class of possible theories beyond the standard model, it appears possible that only using gauge-invariant perturbation theory it will be possible to give accurate predictions. Fortunately, use can be made of results in conventional perturbation theory, and thus requires in many cases quite little effort.

However, for a new method, no matter its grounding in field theory, it is required to explicitly demonstrate its usefulness. As this requires the calculation of genuine non-perturbative quantities, such methods are also necessary to check its predictions. This has been done using lattice methods, but independent checks using different methods, or even experiment \cite{Egger:unpublished}, are desirable. The lattice calculations confirm the predictions of gauge-invariant perturbation theory, and even demonstrate a first candidate where for the case of deviating qualitative predictions the results of gauge-invariant perturbation theory are preferred.

If this evidence could be substantially improved, it appears necessary to reevaluate candidate theories for new physics, whether their low energy degrees of freedom are indeed consistent with the standard model, or not, and whether predictions for new physics are affected. Furthermore, the indications that a scalar lighter than the vector is not possible in the standard model case, up to fermion effects and in direct contradiction to (lattice) perturbation theory, suggests that the check of validity of perturbation theory, no matter if conventional or gauge-invariant, may be called for in more circumstances than expected.

As a final remark, gauge-invariant perturbation theory is limited to theories with a BEH effect. However, the considerations leading to its development, gauge-invariance of physical states, can also raise questions for theories without BEH effect, like technicolor-type theories \cite{Maas:2015gma}. E.\ g., in this case it requires to have gauge-invariant vector particles, essentially light vector mesons, as the lightest degrees of freedom, which is a feature so far not understood.

Summarizing, gauge-invariance seems to require a new set of tools if taken seriously on a fundamental level. This new set of tools is delivered by gauge-invariant perturbation theory \cite{Frohlich:1980gj,Frohlich:1981yi}. It appears that this will change little for the standard model, except in very particular circumstances, but may have fundamental impact beyond the standard model. It seems worthwhile to explore this option, to avoid any possibility to miss standard model background when searching new physics or investing effort in a theory which on conceptual grounds cannot reduce to the standard model at low energies.

\bibliographystyle{bibstyle}
\bibliography{bib}

\end{document}